%% file: autopath.tex
\documentclass{llncs}
\include{commands}

\begin{document}

\setlength{\floatsep}{4pt plus 4pt minus 1pt}
\setlength{\textfloatsep}{4pt plus 2pt minus 2pt}
\setlength{\intextsep}{4pt plus 2pt minus 2pt}
\setlength{\dbltextfloatsep}{3pt plus 2pt minus 1pt}
\setlength{\dblfloatsep}{3pt plus 2pt minus 1pt} 

\setlength{\abovecaptionskip}{3pt}
\setlength{\belowcaptionskip}{2pt}
\setlength{\abovedisplayskip}{2pt plus 1pt minus 1pt}
\setlength{\belowdisplayskip}{2pt plus 1pt minus 1pt}

\title{Similarity Modeling on Heterogeneous Networks via Automatic Path Discovery}
\author{Carl Yang \and Mengxiong Liu \and Frank He \and Xikun Zhang \and Jian Peng \and Jiawei Han}
\authorrunning{Carl Yang et al.}
\institute{University of Illinois at Urbana-Champaign, Urbana, IL 61801, USA \\
\email{\{jiyang3, mliu60, shibihe, xikunz2, jianpeng, hanj\}@illinois.edu}}

\maketitle
\vspace{-10pt}
\input{sec-abstract}
\vspace{-5pt}
\input{sec-intro}
\input{sec-pre}
\input{sec-model}
\input{sec-exp}

\input{sec-con}
\input{sec-ack}

\bibliographystyle{splncs03}
\bibliography{carlyang}
\end{document}

%% file: commands.tex
\usepackage{booktabs} 
\usepackage{amsmath,url,graphicx,amssymb}
\usepackage{amsfonts}
\usepackage{ctable}
\usepackage{multirow}
\usepackage{algorithm}
\usepackage{algpseudocode}
\usepackage{pifont}
\usepackage{color}
\usepackage{subfigure}

\makeatletter
\def\blfootnote{\xdef\@thefnmark{}\@footnotetext}
\makeatother

\newcommand{\mylistbegin}{
  \begin{list}{$\bullet$}
   {
     \setlength{\itemsep}{-2pt}
     \setlength{\leftmargin}{1em}
     \setlength{\labelwidth}{1em}
     \setlength{\labelsep}{0.5em} } }
\newcommand{\mylistend}{
   \end{list}  }

\newcommand{\eg}{\textit{e.g.}}

\newcommand{\ie}{\textit{i.e.}}
\newcommand{\etc}{\textit{etc}}

\newcommand{\wrt}{\textit{w.r.t.~}}
\newcommand{\header}[1]{{\vspace{+1mm}\flushleft \textbf{#1}}}

%% file: sec-abstract.tex
\begin{abstract}
\vspace{-5pt}
Heterogeneous networks are widely used to model real-world semi-structured data. The key challenge of learning over such networks is the modeling of node similarity under both network structures and contents. 
To deal with network structures, most existing works assume a given or enumerable set of meta-paths and then leverage them for the computation of meta-path-based proximities or network embeddings. 
However, expert knowledge for given meta-paths is not always available, and as the length of considered meta-paths increases, the number of possible paths grows exponentially, which makes the path searching process very costly. 
On the other hand, while there are often rich contents around network nodes, they have hardly been leveraged to further improve similarity modeling.
In this work, to properly model node similarity in content-rich heterogeneous networks, we propose to automatically discover useful paths for pairs of nodes under both structural and content information. To this end, we combine continuous reinforcement learning and deep content embedding into a novel semi-supervised joint learning framework. 
Specifically, the supervised reinforcement learning component explores useful paths between a small set of example similar pairs of nodes, while the unsupervised deep embedding component captures node contents and enables inductive learning on the whole network. The two components are jointly trained in a closed loop to mutually enhance each other.
Extensive experiments on three real-world heterogeneous networks demonstrate the supreme advantages of our algorithm.
\end{abstract}
\vspace{-20pt}
\keywords{similarity modeling, heterogeneous networks, deep embedding}
                                                                                                                                                                                                                                                                                                                                                                                                                                                                                                                                                                                                                                                                                                                                                                                                                                                                                                                                                                                                                                                                                                                                                                                                                                                                                                                                                                                                                                                                                                                                                                                                                                                                                                                                                                                                                                                                                                                                                                                                                                                                                                                                                                                                                                                                                                                                                                                                                                                                                                                                                                                                                                                                                                                                                                                                                                                                                                                                                                                                                                                                                                                                                                                                                                                                                                                                                                                                                                                                                                                                                                                                                                                                                                                                                                                                                                                                                                                                                                                                                                                                                                                                                                                                                                                                                                                                                                                                                                                                                                                                                                                                                                                                                                                                                                                                                                                                                                                                                                                                                                                                                                                                                                                                                                                                                                                                                                                                                                                                                                                                                                                                                                                                                                                                                                                                                                                                                                                                                                                                                                                                                                                                                                                                                                                                                                                                                                                                                                                                                                                                                                                                                                                                                                                                                                                                                                                                                                                                                                                                                                                                                                                                                                                                                                                                                                                                                                                                                                                                                                                                                                                                                                                                                                                                                                                                                                                                                                                                                                                                                                                                                                                                                                                                                                                                                                                                                                                                                                                                                                                                                                                                                                                                                                                                                                                                                                                                                                                                                                                                                                                                                                                                                                                                                                                                                                                                                                                                                                                                                                                                                                                                                                                                                                                                                                                  

%% file: sec-intro.tex
\vspace{-5pt}
\section{Introduction}
\label{sec:intro}
\vspace{-5pt}
Networks are commonly used to model relational data such as people with social relations and proteins with biochemical interactions. Recently, increasing research attention has been paid to heterogeneous networks, highlighting multi-typed nodes and connections. Their modeling of rich semantics in terms of both node contents and typed links enables the integration of real-world data from various sources and facilitates wide applications \cite{sun2011pathsim,liu2017semantic,yang2017bridging,yang2018did,zhao2017meta}.

\begin{figure}[h!]
    \centering
        \includegraphics[width=0.85\linewidth]{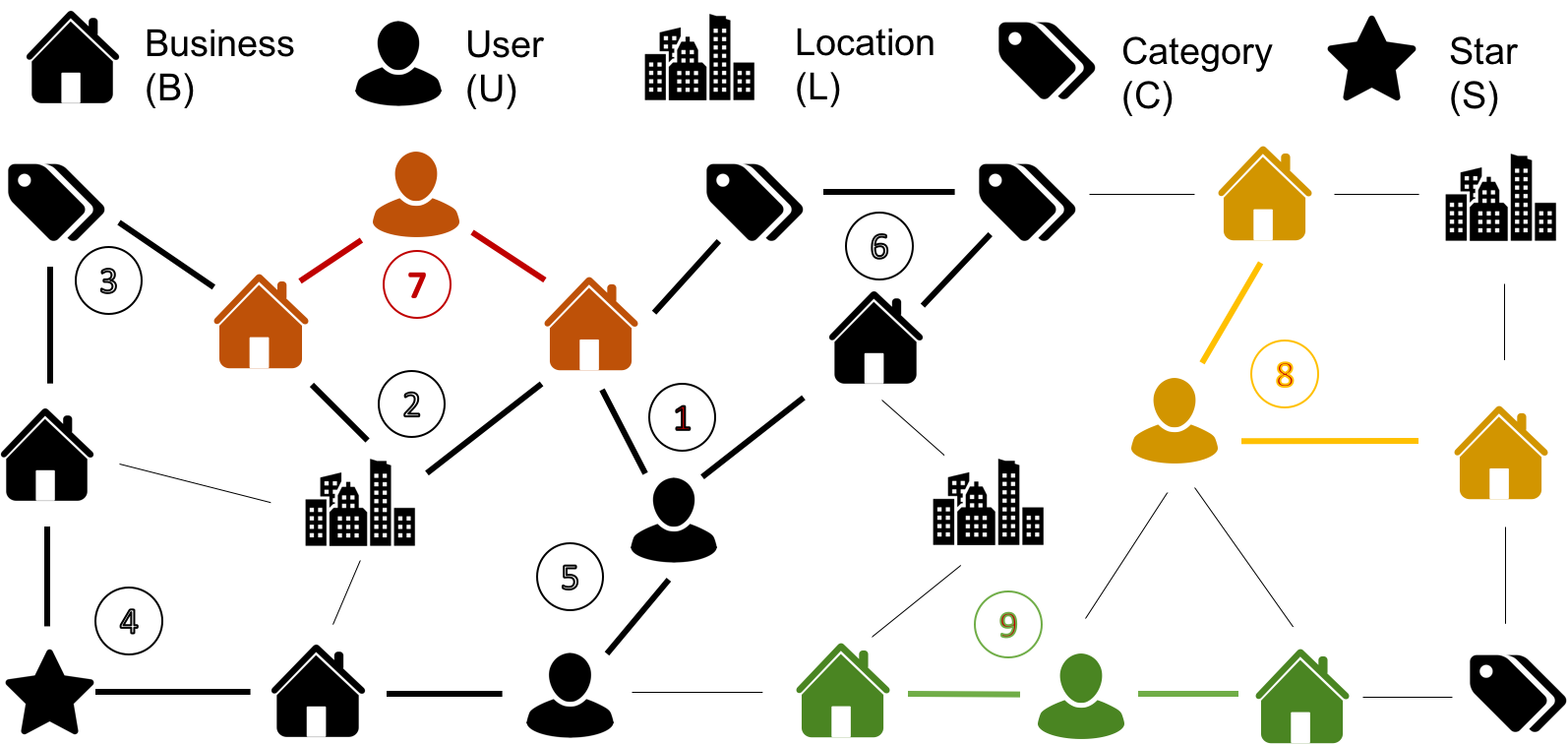}
    \caption{A toy example of modeling the Yelp data with heterogeneous networks.}
    \label{fig:toy}
 \end{figure}

The key challenge of learning with heterogeneous networks is the modeling of node \textit{similarities} (also known as \textit{proximities}) \cite{sun2012mining}. To deal with this, meta-paths have been introduced to constrain the counting of path instances \cite{sun2011pathsim,wang2016relsim} or guide meaningful network embedding \cite{dong2017metapath2vec,shang2016meta}. However, we summarize the drawbacks of most existing heterogeneous network learning algorithms into the following two aspects and explain them in details with our toy example in Figure \ref{fig:toy}.

\header{Drawback 1: Assumption of given or enumerable sets of meta-paths.}
Most existing methods for heterogeneous network modeling assume a known set of useful meta-paths, either given by domain experts or exhaustively enumerable. Then they combine the information of multiple meta-paths through uniform addition \cite{sun2011pathsim,dong2017metapath2vec,shang2016meta,huang2017heterogeneous}, or importance weighing \cite{fu2017hin2vec,zhao2017meta,wang2016relsim,fang2016semantic,meng2015discovering}. However, given any arbitrary heterogeneous network, the process of composing meta-paths according to domain knowledge is ad hoc. Moreover, it is not always efficient or even feasible to enumerate or search for all potentially useful paths, since the number of paths grows exponentially as we consider longer paths, and it is notoriously costly to instantiate the paths on the network.

Consider our toy example in Figure \ref{fig:toy}, which is a simple heterogeneous network constructed with the Yelp data similarly as done in \cite{zhao2017meta}. We only consider five node types: \textsf{businesses} (B), \textsf{users} (U), \textsf{locations} (L), \textsf{categories} (C) and \textsf{stars} (S). 
As for links, we only consider \textsf{users reviewing businesses} (U -- B), \textsf{businesses residing in locations} (B -- L), \textsf{businesses belonging to categories} (B -- C), \textsf{businesses having stars} (B -- S), \textsf{users being friends with users} (U -- U) and \textsf{categories belonging to categories} (C -- C), while other links such as those between adjacent star levels and pairs of geographically nearby locations are ignored for the simplicity of the example.

On this simple heterogeneous network, if we only consider meta-paths between pairs of businesses with length no longer than 4, we already have 6 paths (1-6). Once we increase the length to 5, since meta-paths of length 5 can be composed by two meta-paths of length 3 or one meta-path of length 4 with an additional node, the number of meta-paths of length 5 alone is around 20 ($4\times 4$ for the combination of two paths of length 3 plus $2$ for paths with 3 categories or users in a row between two businesses). 

Note that this is a simplified heterogeneous network with a few node types and link types, and we are only considering meta-paths with lengths no longer than 5, while each meta-path can have millions of instances. Real-world heterogeneous networks can be much more complex. Also, while existing works argue that longer paths are less useful, there exists no solid support for this argument, nor a good way of setting the maximum length of paths to consider.

\header{Drawback 2: No leverage of rich contents around network nodes.}\\
Furthermore, networks can have various contents \cite{yang2017bi}, but no existing algorithm on heterogeneous networks has considered the integration of such rich information. For instance, in the example in Figure \ref{fig:toy}, users have attributes like \textsf{number of reviews}, \textsf{time since joining Yelp}, \textsf{number of fans}, \textsf{average rating of all reviews}. Such contents well characterize user properties like \textit{preference} and \textit{expertise}. 

In this paper, we argue that even instances of the same meta-path can carry rather different semantic meanings. 
To give a few examples, suppose the user on path 7 enjoys high-end restaurants while that on path 8 prefers cheap ones. The two pairs of businesses on the ends of the paths are then close in different ways. Likewise, if the two users on path 7 and 9 have been to very different numbers of places, they may choose the places to go based on quite different criteria, and thus again lead to different path semantics. 
Besides users, categories can be differentiated based on the generality, while locations cover different ranges. Stars also correspond to different similarities, as 1-star means \textsf{equally bad} whereas 5-star means \textsf{comparably fantastic}. 
Due to such observations, existing heterogeneous network learning algorithms are incompetent, because they do not consider the node contents and, as a consequence, model every instance of a meta-path as the same.
It is urgent that we develop a powerful framework to incorporate such semantics and better model node similarity on heterogeneous networks.

\vspace{-5pt}
\header{Insight: Semi-Supervised Learning with limited labeled examples.}\\
In this work, we propose to leverage SSL to capture both structural and content information that is important for measuring the similarities among nodes on heterogeneous networks. 
Given an arbitrary network, unlike existing methods, we do not require a known set of useful meta-paths, nor do we try to enumerate all of them up to a heuristic length limit. 
Instead, we depend on a small number of example pairs of similar nodes which can be easily composed. Then we design an efficient algorithm to automatically explore useful paths on the network under the supervision of these labeled node pairs.
In this way, the structural information on the heterogeneous networks can be fully leveraged.

Moreover, to incorporate content information such as node attributes, we combine an unsupervised objective of content embedding with the supervised path discovery into an SSL framework. By modeling the unlabeled node contents in an unsupervised way, it allows our algorithm to induce the similarity among unlabeled nodes on the whole network, as well as unseen nodes that might be added to the network in the future. It also avoids the requirements for large amounts of training data that cover the whole network.

\vspace{-5pt}
\header{Approach: Reinforcement Learning with Deep Content Embedding.}\\
In this work, we propose \textsc{AutoPath}, to solve the problem of similarity modeling on content-rich heterogeneous networks.

As we discussed before, the number of paths between nodes is exponential to the length. Moreover, searching for paths on networks is notoriously expensive. To deal with such challenges, we leverage reinforcement learning, which has been found efficient in sequential decision making and successfully applied for path exploration on knowledge bases \cite{xiong2017deeppath,das2017go}. However, to the best of our knowledge, there is no previous work on employing reinforcement learning to model heterogeneous networks, which have quite a few unique properties, such as the large action spaces at each node when growing the paths and the large numbers of valid paths between each pair of nodes. Such properties make the direct application of existing algorithms on knowledge bases to heterogeneous networks impossible.
Another major distinction between heterogeneous networks and knowledge bases is the prevalence of rich node contents, which has hardly been explored before by existing algorithms.
The existence of such node contents that potentially differentiate the semantics on instances of the same meta-paths further increases the difficulty of similarity modeling over heterogeneous networks.
Such situations, as we will discuss more in Section \ref{sec:pre}, urge the development of a specifically designed reinforcement learning framework. 

To overcome the challenges of large action spaces and node contents simultaneously, we leverage continuous reinforcement learning and incorporate deep content embedding to learn the state representations. 
Specifically, continuous policy gradient effectively estimates similar actions and avoids the explicit search over all discrete actions. Moreover, we devise conjugate deep autoencoders to capture node types and contents, and jointly train them with the policy and value networks of the reinforcement learning agent in a closed loop, so as to allow the mutual enhancement between embedding and learning. More details of our models are discussed in Section \ref{sec:model}.

As we will demonstrate in Section \ref{sec:exp}, our proposed \textsc{AutoPath} algorithm is able to break free the requirements of known sets of meta-paths, leverage node contents, and achieve state-of-the-art performance on the task of similarity search with very limited supervision. Extensive quantitative experiments and qualitative analysis on three real-world heterogeneous networks demonstrate the advantages of \textsc{AutoPath} over various state-of-the-art heterogeneous network modeling algorithms. 


%% file: sec-pre.tex
\vspace{-5pt}
\section{Preliminaries}
\label{sec:pre}
In this section, we briefly introduce the key concepts and relevant techniques of heterogeneous network modeling and reinforcement learning. Due to space limit, a broader discussion of related works is placed into our \textit{Supplementary Materials}.

\vspace{-5pt}
\subsection{Heterogeneous Network Modeling}
Heterogeneous network has been intensively studied due to its power of accommodating multi-typed interconnected data \cite{sun2012mining,sun2011pathsim,dong2017metapath2vec,yang2017bridging}. 
In this work, we stress that rich contents are prevalently available on nodes in the networks, and we define content-rich heterogeneous networks as follows.
\begin{definition}{Content-Rich Heterogeneous Network.} 
A content-rich heterogeneous network is defined as a directed graph $\mathcal{N}=\{\mathcal{V}, \mathcal{E}, \mathcal{A}\}$. For each node $v \in \mathcal{V}$ and its corresponding node type $\phi(v)=T$, a content vector $A^T_v \in \mathcal{A}$ is associated with $v$. Depending on the node type $T$ and available data, $A^T$ can be categorical, numerical, textual, visual, \etc., or any mixture of them.
\end{definition}

To properly model heterogeneous networks, \cite{sun2011pathsim} introduces the concept of meta-path, which has been the golden measure of similarity among nodes on heterogeneous networks \cite{sun2011pathsim,wang2016relsim,fang2016semantic,meng2015discovering,shi2017prep,wang2015knowsim}, and recently have also enabled various heterogeneous network embedding algorithms \cite{dong2017metapath2vec,shang2016meta,huang2017heterogeneous,fu2017hin2vec,shi2018aspem,wan2015graph}.
However, most existing heterogeneous network modeling algorithms assume a given or enumerable set of useful meta-paths up to a certain empirically decided length, which is not always practical. Moreover, they do not consider contents in the networks, and thus regard all instances of the same meta-paths as the same.

\subsection{Reinforcement Learning}
The main challenge of heterogeneous network modeling without a known set of meta-paths is to automatically explore and find the useful ones, which is naturally a combinatorial problem.
For automatic path discovery on heterogeneous networks, as we consider $K$ types of nodes and meta-paths of length $L$, the number of all possible meta-paths can be at the same scale as $K^L$. Moreover, we stress that on content-rich heterogeneous networks, instances of the same meta-paths can carry different semantics, and the search space is further enlarged to approximately $\rho^L$, where $\rho$ is the average out-degree of nodes on the network and is often much larger than $K$. 

Reinforcement learning has been intensively studied for solving complex planning problems with consecutive decision makings, such as robot control and human-computer games \cite{mnih2015human,sutton2000policy}. 
Recently, there are several approaches based on reinforcement learning to tackle the combinatorial optimization problems over network data \cite{bello2016neural,khalil2017learning}, as well as reasoning over knowledge bases \cite{das2017go,xiong2017deeppath}, which are shown to be effective. Motivated by their success, we aim to leverage reinforcement learning to efficiently solve the combinatorial problem of automatic path discovery on heterogeneous networks. 

Different from knowledge bases, although content-rich heterogeneous networks have fewer node types, each type has much larger number of nodes. Categorical actor networks used in \cite{das2017go,xiong2017deeppath} have poor convergence property in our heterogeneous network setting. To address this issue, continuous reinforcement learning serves as an appropriate paradigm. Our action is applied in the deep embedding space which is trained together with conjugate autoencoders to represent node types and contents. Unlike DDPG \cite{lillicrap2015continuous} or Q-learning \cite{mnih2015human} which learn a deterministic policy, our algorithm is designed to learn a probability distribution over actions as a policy. By sampling from the learned policy, our framework assigns large probabilities to high-quality paths.  To briefly summarize, our algorithm leverages both structural and content information and automatically discover meaningful paths, under the guidance of limited labeled data.

%% file: sec-model.tex
\section{\textsc{AutoPath}}
\label{sec:model}
In this section, we describe our \textsc{AutoPath} algorithm, which combines reinforcement learning and deep embedding over content-rich heterogeneous networks into a semi-supervised learning framework.

\subsection{Overall Semi-Supervised Learning Framework}
We start with a formal definition of our problem.
\begin{definition}{Similarity Modeling.}
Consider a content-rich heterogeneous network $\mathcal{G}=\{\mathcal{V}, \mathcal{E}, \mathcal{A}\}$ with a corresponding type function $\phi$. 
The problem of similarity modeling is to measure the similarity between any pair of nodes, under the consideration of various meta-paths and rich node contents on the path instances.
\end{definition}
We stress that similarity modeling is the key challenge of learning with content-rich heterogeneous networks, as its solution naturally enables various subsequent tasks like link prediction, node classification, community detection and so on.

\begin{figure}[h!]
\vspace{10pt}
    \centering
        \includegraphics[width=1\linewidth]{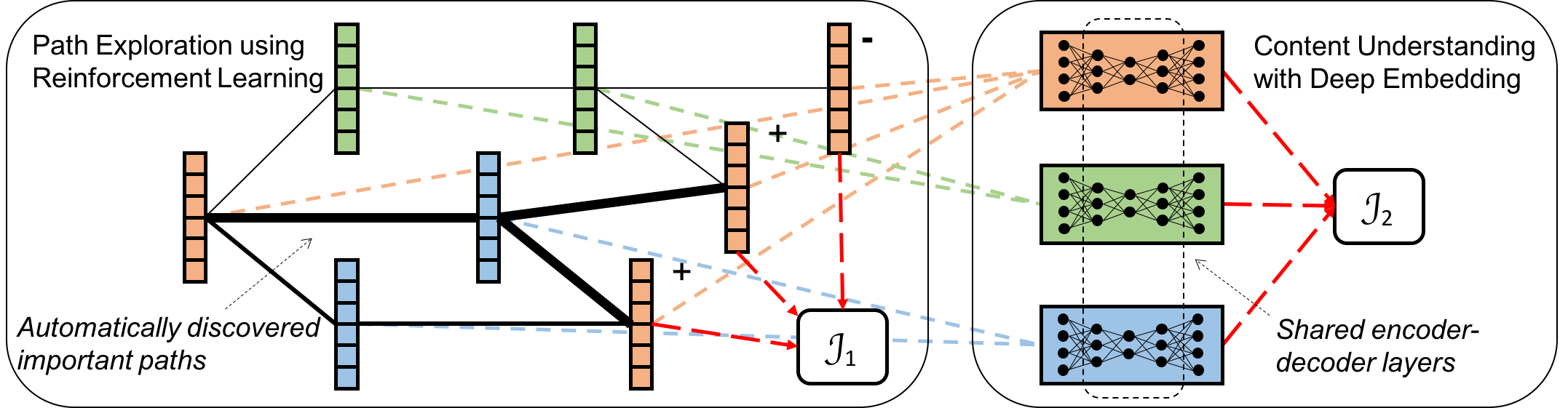}
    \caption{Overall framework of \textsc{AutoPath}: Nodes are encoded by their embedding vectors based on both structure and content information on the network, where different colors denote different node types. The black solid lines denote actual network links, and the line weights denote the importance of the paths discovered by the algorithm. The colored dash lines indicate the connections between node embeddings and their content embedding models \wrt~the corresponding node types. The supervised module with an objective $\mathcal{J}_1$ is trained \wrt~the labeled node pairs in the given example set, and the unsupervised module with an objective $\mathcal{J}_2$ is trained over the whole network.}
    \label{fig:model}
\end{figure}

In this work, we aim to automatically learn the important meta-paths and node contents by leveraging limited labeled data. Therefore, besides a graph $\mathcal{G}=\{\mathcal{V}, \mathcal{E}, \mathcal{A}\}$, we consider the basic input as a set of example similar pairs of nodes $\mathcal{P}$, upon which we build a supervised learning module using reinforcement learning to explore their prominent connecting paths characterized by network links $\mathcal{E}$. To make the learning algorithm efficient and aware of node contents $\mathcal{A}$, we further build an unsupervised learning module with deep content embedding, which also enables inductive learning on the whole network $\mathcal{G}$ not necessarily covered by $\mathcal{P}$. Figure \ref{fig:model} shows the overall framework of \textsc{AutoPath}, and in what follows, we describe the two major components of this framework in details.

\subsection{Path Exploration using Reinforcement Learning}
\header{Learning Paradigm.}
As we discussed before, automatic path exploration is essentially a combinatorial problem over enormous search spaces, which cannot be well solved by exhaustive enumeration or searching with greedy pruning. Motivated by the recent success of reinforcement learning on sequential decision making, we propose to leverage the following paradigm for efficient path exploration.

For an example pair of similar nodes $p=\{s, t\}\in \mathcal{P}$, we call $s$ a \textit{start node} and $t$ a \textit{target node}. From each start node, we repeatedly train the reinforcement learning agent by looking for the next node to go on the network. A partial solution is represented as a sequence ${S}=(s, v_1, v_2, \ldots)$. At each step, based on the model parameters and the current state, the agent will either choose a neighboring node to go to ($v_k$) or return to the start node ($s$).
Every time the agent reaches the target node ($t$), it gets a positive reward and returns to the start node ($s$). We will depict the details of $\mathcal{P}$ on different datasets in Section \ref{sec:exp}.

\header{Framework Representation.}
To deal with the aforementioned large action space challenge, we propose to leverage a novel network embedding method, which captures the node types and contents into a low-dimensional latent space. The details of the embedding method are deferred to the next subsection. 
Similar to \cite{bello2016neural}, our neural network architecture models a stochastic policy $\pi(a \mid {S}, \mathcal{G})$, where $a$ is the action of selecting the next node from the network $\mathcal{G}$ and $S$ is the current partial solution. 


We define the components in our reinforcement learning framework as follows.
\begin{enumerate}
\item \textbf{State:} A state ${S}$  is a sequence of nodes we have selected. Based on our novel network embedding method, a state is represented by a $\kappa$-dimensional vector $\sum_{v \in S} \mathbf{x}_v$, while it is also possible to use mean pooling, max pooling or neural networks like LSTM.
\item \textbf{Action:} An action $a$ is a node $v \in \mathcal{V}$. We cast the details of actions later.
\item \textbf{Reward:} The reward $r$ of taking action $a$ at state $S$ is $r=1$ if $a=t$, and $r=0$ otherwise.
\item \textbf{Transition:} The transition is deterministic by simply adding the node $v$ we have selected according to action $a$ to the current state ${S}$. Thus, the next state ${S}':=({S},v)$.
\end{enumerate}

Our actor network (policy network) $\mu_\Theta({S})$ and critic network (value function) $\nu_\Theta({S})$ are both fully connected feedforward neural networks, each containing four layers including two hidden layers of size $H$, as well as the input and output layers. 
Rectified linear unit (ReLU) is used as the activation function for each layer, and the first hidden layer is shared between two networks. Both networks' inputs are $\kappa$-dimensional node embeddings. The output of the actor network $\mu_\Theta({S})$ are $\kappa$-dimensional vectors $ \mu$ and $\sigma^2$, whereas the output of critic network $\nu_\Theta({S})$ is a real number.

\header{Learning Algorithm.}
To overcome the large action space problem, we adopt continuous policy gradient as our learning algorithm. Our policy selects actions in node embedding space \cite{schulman2015trust,lillicrap2015continuous}. At each time step, we select a continuous vector and then retrieve the closest node from the current neighborhood plus the start node by comparing the action vector with the node embeddings. 

Consider our policy $\pi(a \mid {S})$, unlike in the discrete action domain where the action output is a \textit{softmax} function, here the two outputs of the policy network are two real number vectors which we treat as the mean $\mu$ and  variance $\sigma^2$ of a multi-dimensional normal distribution with a spherical covariance $\Sigma=\sigma^2 I$. To act, the input is passed through the model to the output layer where a Gaussian exploration is determined by $\mu$ and $\sigma^2$ as
\begin{equation}
\pi(a \mid {S}, \{\mu, \Sigma \})=\frac{1}{\sqrt{2\pi \left| \Sigma \right|}} \exp \left(-\frac{1}{2}(a-\mu)^T \Sigma^{-1} (a-\mu)\right).
\end{equation}



Since our goal is to find the important path ${S}$, our training loss is 
\begin{equation}
\mathcal{J}_p(\Theta \mid \mathcal{G}) = -\mathbb{E}_{\tau \sim p_\Theta({S}\mid \mathcal{G})} R(\tau),
\label{obj_pg}
\end{equation}
where $\tau$ denotes an episode of the state-action trajectory, $\Theta$ is the set of parameters, and $R$ is the reward. $\mathcal{J}_p$ is called the surrogate loss in reinforcement learning which evaluates the quality of the entire path $S$ constructed by $\tau$. To derive the gradient of $\mathcal{J}_p$, we use the policy gradient theorem \cite{sutton2000policy} which gives
\begin{align}
\nabla_\Theta \mathcal{J}_p&=-\frac{1}{\alpha} \sum_{i=1}^{\alpha} \sum_{t=0}^{T-1} \nabla_\Theta \pi_\Theta (a_t^{(i)}\mid S_t^{(i)}) \hat{A}_t,\\
\hat{A}_t &= \left(\sum_{k=t}^{T-1}r(S_k^{(i)}, a_{k}^{(i)}) -b(S_k^{(i)})\right),
\end{align}
where $\alpha$ is the number of trajectories, $T$ is the trajectory length, $\hat{A}_t$ is advantage and $b$ is the baseline for variance reduction. By exploiting the fact that

\begin{equation}
\nabla_\Theta \pi_\Theta(a \mid S) = \pi_\Theta (a \mid S) \frac{\nabla_\Theta \pi_\Theta(a \mid S) }{\pi_\Theta(a \mid S)}=\pi_\Theta (a \mid S) \nabla_\Theta \log \pi_\Theta(a \mid S),
\end{equation}
we have the approximate gradient estimator as
\begin{equation}
g=\mathbb{E}_t[\nabla_\Theta \log \pi_\Theta (a_t \mid S_t) \hat{A}_t],
\end{equation}
where $\mathbb{E}_t$ denotes the empirical average over a mini-batch of samples in the algorithm that alternates between sampling and optimization using policy gradient.




In order to reduce the variance, we choose the value function $\mathcal{V}_\Theta$ as the baseline. $\mathcal{V}_\Theta$ is learned by using Monte Carlo method to minimize the loss 
\begin{align}
\mathcal{J}_v=\|\mathcal{V}_\Theta(S_t) - \sum_{k=t}^{T-1}r(S_k, a_k) \|_2^2.
\end{align} 

Subsequently, we define our policy gradient loss as the sum of surrogate loss and value function loss, \ie, $\mathcal{J}_1 = \mathcal{J}_p + \mathcal{J}_v$, which can be regarded as a supervised loss under the example similar pairs of nodes.


\subsection{Content Understanding with Deep Embedding}
\header{Conjugate Autoencoders.}
In order to make \textsc{AutoPath} aware of node contents and able to perform inductive learning on the whole network, we design a novel unsupervised node embedding method. Unlike existing network embedding methods designed to capture link structures, we aim to represent node types and contents in a shared low-dimensional space. To this end, we get inspired by recent success in deep learning for feature composition \cite{le2013building}, which has been proven advantageous in capturing intrinsic features within complex contents in an unsupervised learning fashion.

To be specific, we propose conjugate autoencoders, which is a novel variant of deep denoise autoencoder. It consists of two non-linear feedforward neural network layers, \ie, two encoder layers and two decoder layers. The first encoder layers and the last decoder layers have individual embedding weights for each node type, while the other two layers are shared across different node types, as demonstrated in Figure \ref{fig:model}. Therefore, the embedding $\mathbf{x}_i$ for node $v_i$ of type $k$ (\ie, $\phi(v_i)=k$) is computed as
\begin{align}
\mathbf{x}_i = \mathbf{f}_e^o(\mathbf{f}_e^k(\mathbf{a}_i)), \text{ where } \mathbf{f}_e^j(\mathbf{x})=ReLU(\mathbf{W}_e^j Dropout(\mathbf{x})+\mathbf{b}_e^j).
\end{align}
Similarly, the reconstructed feature $\mathbf{\tilde{a}}_i$ of node $v_i$ is computed as
\begin{align}
\mathbf{\tilde{a}}_i= \mathbf{f}_d^k(\mathbf{f}_d^o(\mathbf{x}_i)), \text{ where } \mathbf{f}_d^j(\mathbf{x})=ReLU(\mathbf{W}_d^j Dropout(\mathbf{x})+\mathbf{b}_d^j).
\end{align}
The parameters in $\mathbf{f}_e^o$ and $\mathbf{f}_d^o$ are shared across all node types, while the parameters in $\{\mathbf{f}_e^k, \mathbf{f}_d^k\}_{k=1}^K$ are different for each node type. 

\header{Content Reconstruction Loss.}
To learn the intrinsic node features in an unsupervised fashion, a \textit{content reconstruction loss} is computed over the whole network as
\begin{align}
\mathcal{J}_r=\sum_{i=1}^n l(\mathbf{a}_i,\mathbf{\tilde{a}}_i).
\end{align}
Depending on the contents in the datasets, $l$ can be implemented either as a cross entropy (for binary features, such as user attributes) or a mean squared error (for continuous features, such as TF-IDF scores of words). 

\header{Type Discrimination Loss.}
While $\mathcal{J}_r$ enforces the capture of node contents, node embeddings computed in this way does not necessarily discriminate different types of nodes in the shared embedding space, which weakens the ability of the algorithm to differentiate various meta-paths. To deal with this, we further impose a \textit{type discrimination loss} over the whole network as
\begin{align}
\mathcal{J}_d= -\sum_{i=1}^n log(p(i)), \text{ where } p(i)=\frac{exp(\mathbf{W}^{\phi(v_i)}_c\mathbf{x}_i)}{\sum_k exp(\mathbf{W}^k_c\mathbf{x}_i)}.
\end{align}
It is basically a softmax classifier towards node types with cross-entropy loss, which acts as adversarial to the shared reconstruction loss to make sure different types of nodes do not mingle too much in the shared embedding space. 

The two losses can be combined with a tunable weighting parameter $\lambda$ as $\mathcal{J}_2=\mathcal{J}_r+\lambda\mathcal{J}_d$. We use $\Phi$ to denote all parameters related to these two losses.

\subsection{Joint Training of Reinforcement Learning and Deep Embedding}
\header{Training Pipeline.}
To realize our SSL framework, we integrate the training of reinforcement learning and deep embedding into a joint learning pipeline, with the overall loss $\mathcal{J} = \mathcal{J}_1+\mathcal{J}_2$.
We firstly pre-train the content embedding with all parameters in $\Phi$ until $\mathcal{J}_2$ is sufficiently small, which captures the intrinsic distribution of node contents in a low-dimensional space. Then we detach the encoder layers and learn the rest of the model through co-training. Such detachment and separation of pre-training and co-training are necessary for allowing the node embeddings to become different for nodes even with the same contents to respect the network structures. 
Specifically, during co-training, we iteratively train the actor and critic networks by updating the parameters in $\Theta$, and the embedding networks by updating the parameters in $\Phi$ except for those in the encoder. Note that, in both processes, the node embeddings $\mathcal{X}$ will also get updated, to reflect both important network structures and node contents. 
In each epoch, when updating $\Theta$ and $\mathcal{X}$, we sample a set $\Omega$ of $\alpha$ trajectories of length $m$ using the current policy $\pi_\Theta(a \mid \mathcal{S})$, with each trajectory starting from a random start node in the set of example node pairs $\mathcal{P}$, and construct the surrogate loss and value function loss in $\mathcal{J}_1$; when updating $\Phi$, we sample a set $\Psi$ of $\beta$ nodes from all nodes $\mathcal{V}$ in the whole network $\mathcal{G}$, and compute the reconstruction loss and discrimination loss in $\mathcal{J}_2$. Mini-batch SGD is then used to optimize the objectives iteratively for $\gamma$ epochs, where all model parameters in $\{\Theta, \Phi, \mathcal{X}\}$ are updated by Adam \cite{kingma2014adam}. 
We released our code with a demo function on Github\footnote{https://github.com/yangji9181/AutoPath} and also included it in our \textit{Supplementary Materials}.

\header{Computational Complexity.}
We theoretically analyze the complexity of \textsc{AutoPath}. 
For the reinforcement learning component, during each step of training, \textsc{AutoPath} generates a target mean $\mu_\Theta$ in constant time and then selects a node from $\mathcal{G}$ that is the closest to $\mu_\Theta$. Note that, to grow a path, we only need to compare nodes in the direct neighborhood of the current node plus the start node, the size of which is much smaller than $n$ and can be regarded as a constant number $\rho$.
Since computing the quality function and updating the neural network model based on particular trajectories take constant time, the overall complexity of training and planning with the reinforcement learning agent is $O(\alpha \rho m)$ in each epoch.
For the deep embedding component, \textsc{AutoPath} uniformly samples the nodes in $O(\beta)$ time, and then compute the losses and update the models in $O(1)$ time. Therefore, the overall training time of \textsc{AutoPath} is $O((\alpha \rho m+\beta)\gamma)$.
The time of model inference for particular nodes is ignorable compared with model training.

%% file: sec-exp.tex
\section {Experimental Evaluations} 
\label{sec:exp}
In this section, we evaluate the performance of our proposed \textsc{AutoPath} algorithm on three real-world content-rich heterogeneous networks in different domains, \ie, IMDb from a movie rating platform\footnote{http://www.imdb.com/}, DBLP from an academic publication collection\footnote{https://dblp.uni-trier.de/}, and Yelp from a business review service\footnote{https://www.yelp.com/}.
Through extensive quantitative experiments and qualitative analysis in comparison with various baselines, we show that \textsc{AutoPath} can efficiently leverage both structural and content information on heterogeneous networks, which leads to supreme performance on the key task of similarity modeling.

\subsection{Experimental Settings}
\vspace{-5pt}
\header{Datasets.} We describe the datasets we use as follows and the statistics are summarized in Table \ref{tab:stat}.
\begin{enumerate}
\item \textbf{IMDb}: We use the MovieLens-100K dataset\footnote{https://grouplens.org/datasets/movielens/100k/} made public by \cite{harper2016movielens}. 
There are four types of nodes in the network, \ie, \textsf{users} (U), \textsf{movies} (M), \textsf{actors} (A), and \textsf{directors} (D). The edge types include \textsf{users reviewing movies}, \textsf{actors featuring in movies}, and \textsf{director making movies}. The contents we use for users include simple demographics like \textsf{age, gender, occupation, zipcode}. For movies, actors and directors, we collect the first textual paragraph of the main content in their corresponding Wikipedia\footnote{https://en.wikipedia.org/wiki/Main\_Page} page if available.

\item \textbf{DBLP}: We use the Arnetminer dataset V8\footnote{https://aminer.org/citation} collected by \cite{tang2008arnetminer}.
It contains four types of nodes, \ie, \textsf{authors} (A), \textsf{papers} (P), \textsf{venues} (V), and \textsf{years} (Y). The edge types include \textsf{authors writing papers}, \textsf{papers citing papers}, \textsf{papers published in venues}, and \textsf{papers published in years}. As for contents, we use titles and abstracts for papers, full names for venues, and also the first textual paragraph of the main content in Wikipedia for authors if available.

\item \textbf{Yelp}: We use the public dataset from the Yelp Challenge Round 11\footnote{https://www.yelp.com/dataset}. Following an existing work that models Yelp data with heterogeneous networks \cite{zhao2017meta}, we extract five types of nodes, \ie, \textsf{businesses} (B), \textsf{users} (U), \textsf{locations} (L), \textit{categories} (C), and \textsf{stars} (S). The edge types include \textsf{users reviewing businesses}, \textsf{businesses belonging to categories}, \textsf{businesses residing in locations}, \textsf{businesses having average stars}, \textsf{category related to categories} and \textsf{users being friends with users}. We further extract contents for businesses like \textsf{latitudes, longitudes, review counts}, \etc., and for users like \textsf{review counts, time since joining Yelp, number of fans, average stars}, \etc. For nodes with no additional contents but a name like categories (\eg, \textsf{Mexican, Burgers, Gastropubs}) and locations (\eg, \textsf{San Francisco, Chicago, London}), we use the pre-trained word embeddings\footnote{https://nlp.stanford.edu/projects/glove/} provided by \cite{pennington2014glove} as initial contents. 

\end{enumerate}
\vspace{-5pt}
As we can see, the structures and sizes of networks are quite different across the experimented datasets, and the network contents are of various types including categorical, numerical, textual and mixtures of them. In this work, we model all textual contents simply as bag-of-words.

\begin{table}[h!]
\centering
  \vspace{10pt}
 \begin{tabular}{|c||c|c|c|c|c|c|}
   \hline
{\bf Dataset}&{\bf Size}&{\bf \#Types}&{\bf \#Nodes}&{\bf \#Links}&{\bf \#Classes}&{\bf \#Pairs}\\
  \hline
  \hline
{\bf IMDb}& 16.1MB & 4 & 45,913 & 153,645 & 23 & 4,000 \\
\hline
{\bf DBLP}& 4.33GB & 4 & 335,185 & 2,704,655 & 4 & 10,000 \\
\hline
{\bf Yelp}& 6.52GB & 5 & 1,123,649 & 8,912,736 & 6 & 20,000\\
\hline
 \end{tabular}
  \vspace{10pt}
 \caption{ \label{tab:stat}Statistics of the three experimented public datasets.}
 \vspace{-10pt}
\end{table}

\vspace{-5pt}
\header{Baselines.}
We compare with both path matching and network embedding based heterogeneous network modeling algorithms to comprehensively evaluate the performance of \textsc{AutoPath}.\\
$\bullet$ \textbf{PathSim} \cite{sun2011pathsim}: Normalized meta-path constrained path counts for measuring node similarity on heterogeneous networks.\\
$\bullet$ \textbf{RelSim} \cite{wang2016relsim}: Exhaustive meta-path enumeration up to a given length and supervised weighting for combining the normalized counts of multiple meta-paths.\\
$\bullet$ \textbf{FSPG} \cite{meng2015discovering}: Greedy meta-path search to a given length and similarity computation through a linear combination of biased path constrained random walks.\\
$\bullet$ \textbf{PTE} \cite{tang2015pte}: Heterogeneous network embedding by decomposing the network into a set of bipartite networks and capturing first and second order proximities.\\
$\bullet$ \textbf{Metapath2vec} \cite{dong2017metapath2vec}: Heterogeneous network embedding through heterogeneous random walks and negative sampling.\\
$\bullet$ \textbf{ESim} \cite{shang2016meta}: Heterogeneous network embedding through meta-path guided path sampling and noise-contrastive estimation. \\

\vspace{-20pt}
\header{Evaluation protocols.}
We study the efficacy of all algorithms on similarity modeling, which can be naturally evaluated under the setting of standard link prediction. The links are generated from additional labels of semantic classes not directly captured by the networks.
For IMDb, we use all 23 available genres such as \textsf{drama, comedy, romance, thriller, crime} and \textsf{action}.
For DBLP, we use the manual labels of authors from four research areas, \ie, \textsf{database, data mining, machine learning} and \textsf{information retrieval} provided by \cite{sun2011pathsim}. 
For Yelp, we extract six sets of businesses based on some available attributes, \ie, \textsf{good for kids, take out, outdoor seating, good for groups, delivery} and \textsf{reservation}. 
For each dataset, we assume that movies (businesses, authors) within each semantic class are similar in certain ways, and generate pairwise links among them.

Following the common practice in \cite{fang2016semantic,meng2015discovering}, we firstly sample certain amounts of linked pairs of nodes, the numbers of which are listed in Table \ref{tab:stat}. 
We use them as training data, \ie, example pairs of similar nodes. 
Since all pairs are positive, we also randomly generate an equal amount of negative pairs, each consisting of two entities not in the same semantic class.
\textit{PathSim} needs no training, while \textit{RelSim} and \textit{FSPG} are both trained on the training data in a supervised way. 
For embedding algorithms, we compute the embeddings in an unsupervised way on the whole network, and train a standard SVM\footnote{http://scikit-learn.org/stable/modules/svm.html} on the training data. 
For \textsc{AutoPath}, we train the reinforcement learning agent with the training data and deep embedding on the whole network. After training, similarity scores can be computed by starting from any particular node, planning with the agent for multiple times, and taking the empirical probabilities of reaching the target nodes. 
For testing, we randomly select $10\%$ start nodes disjointly with the training pairs, and retrieve all target nodes from the same semantic class for each of them to form the ground-truth lists.
Each baseline ranks all nodes on the network \wrt~each start node, and we compute the average \textit{precision at $K$}, \textit{recall at $K$} and \textit{AUC} over all selected start nodes, which are the standard evaluation metrics for link prediction \cite{han2011data}.
We also record the runtimes of all algorithms. 

\header{Parameter settings.}
When comparing \textsf{AutoPath} with the baseline methods, we slightly tune the parameters via cross-validation. For the IMDb dataset, the parameters are empirically set to the following values: 
For reinforcement learning, we set the length of trajectories $m$ to $10$, the sample size $\alpha$ to 400;
for deep embedding, we set the sample size $\beta$ to 2000 and the weighting factor $\lambda$ to 0.1;
for both components, we set the size of hidden layers to 64, and the number of epochs $\gamma$ to 200.
The parameters on other datasets are slightly different due to different data sizes.
During cross-validation, we find \textsf{AutoPath} to be quite robust across different parameter settings. 
All parameters of the compared baselines are either set as given in the original work on the same datasets, or tuned to the best through standard five-fold cross validation on each dataset.

\subsection{Quantitative Evaluation} 
\begin{figure}[h!]
\centering
\subfigure[Precision-IMDb]{
\includegraphics[width=0.33\textwidth]{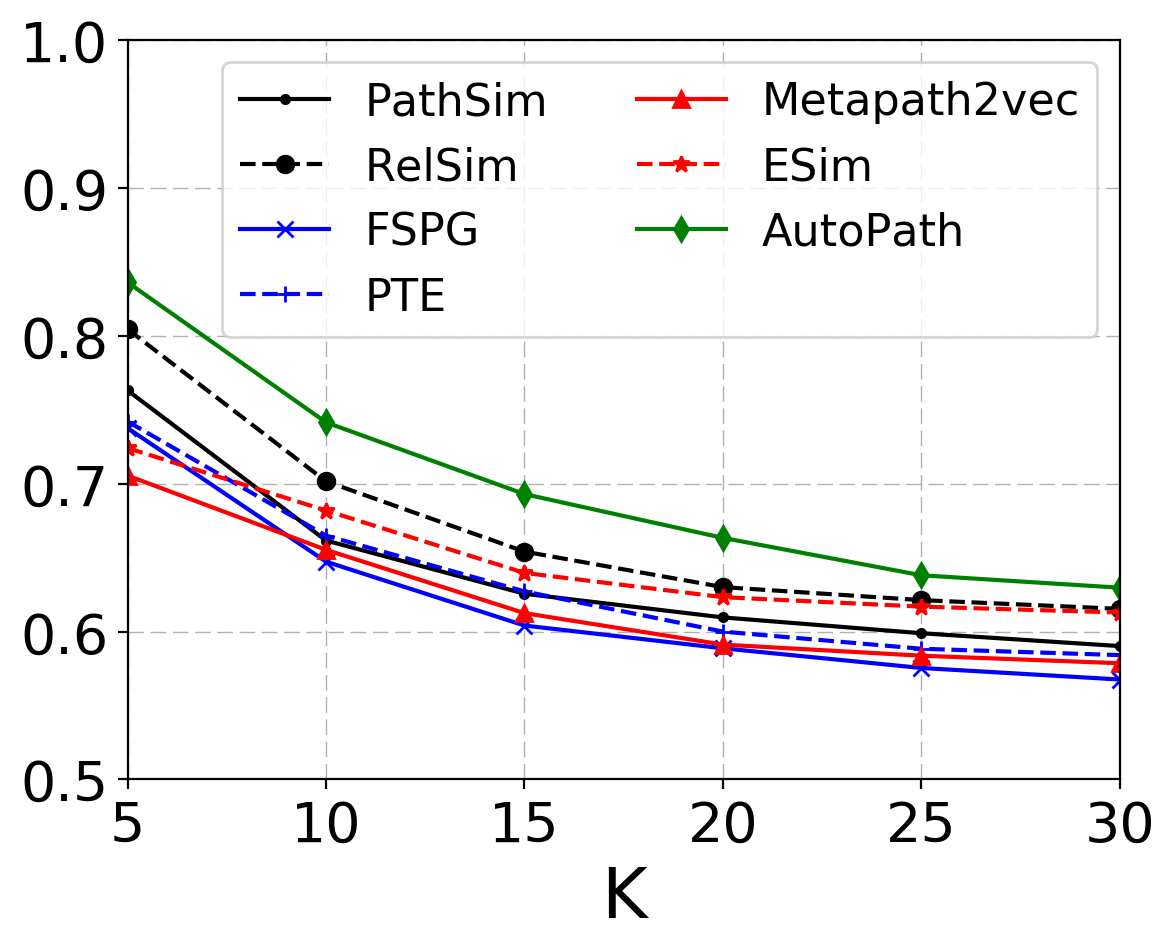}}
\hspace{-10pt}
\subfigure[Precision-DBLP]{
\includegraphics[width=0.33\textwidth]{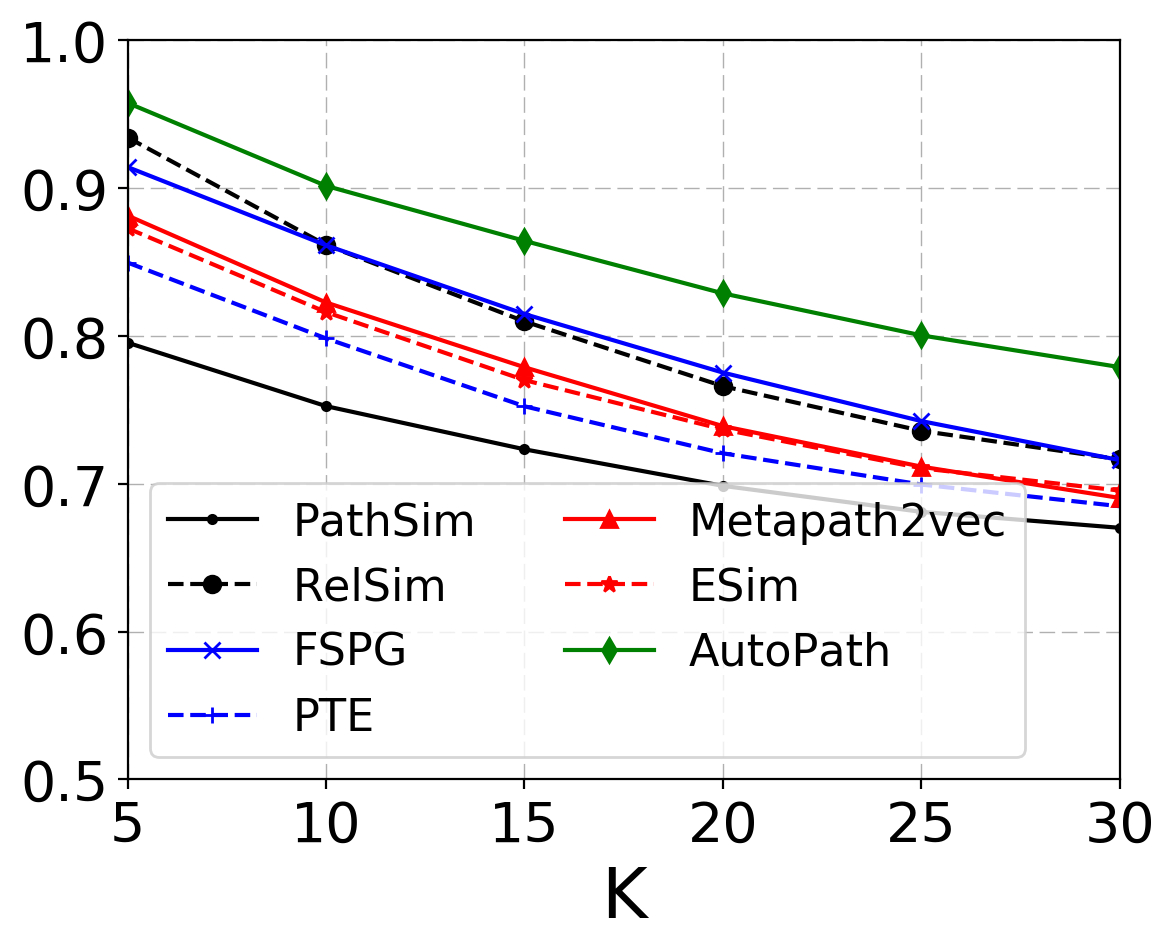}}
\hspace{-10pt}
\subfigure[Precision-Yelp]{
\includegraphics[width=0.33\textwidth]{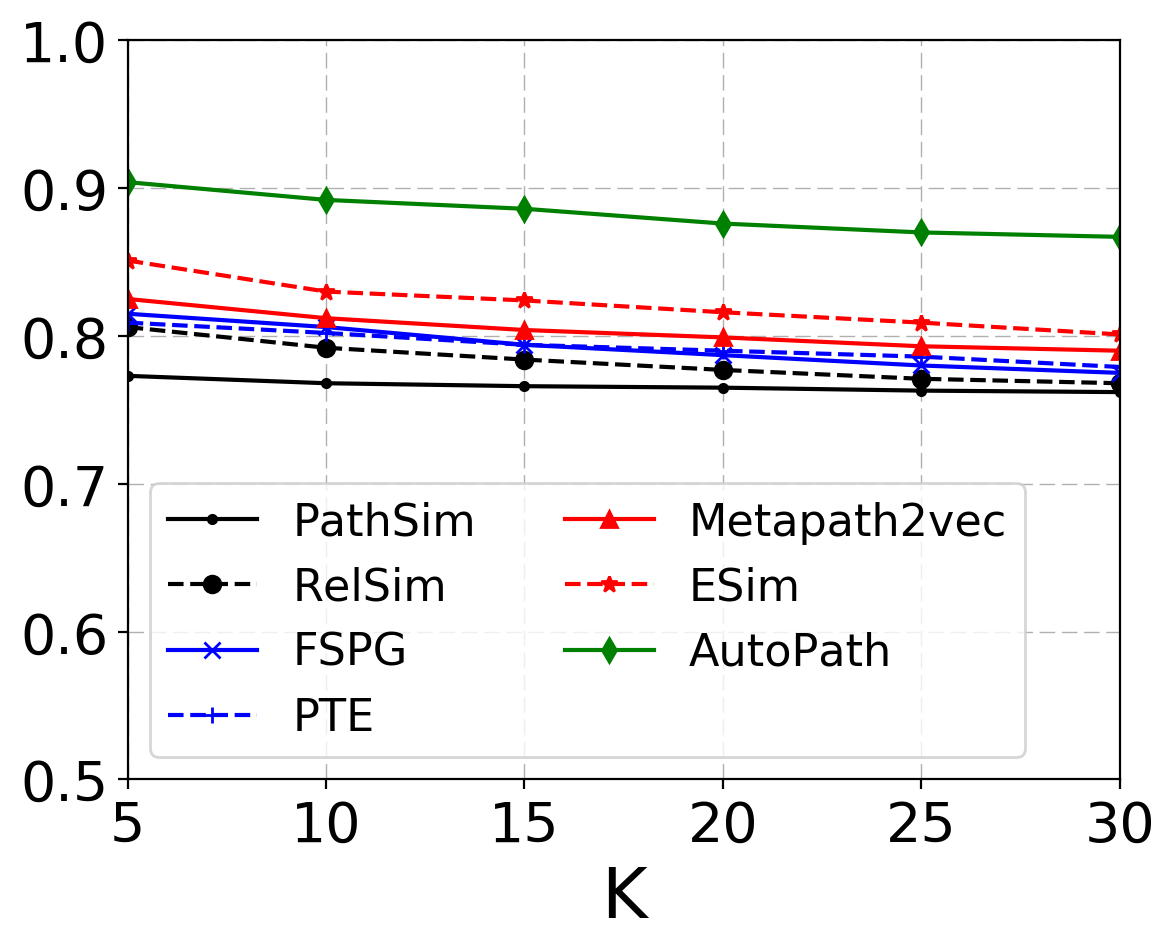}}

\subfigure[Recall-IMDb]{
\includegraphics[width=0.33\textwidth]{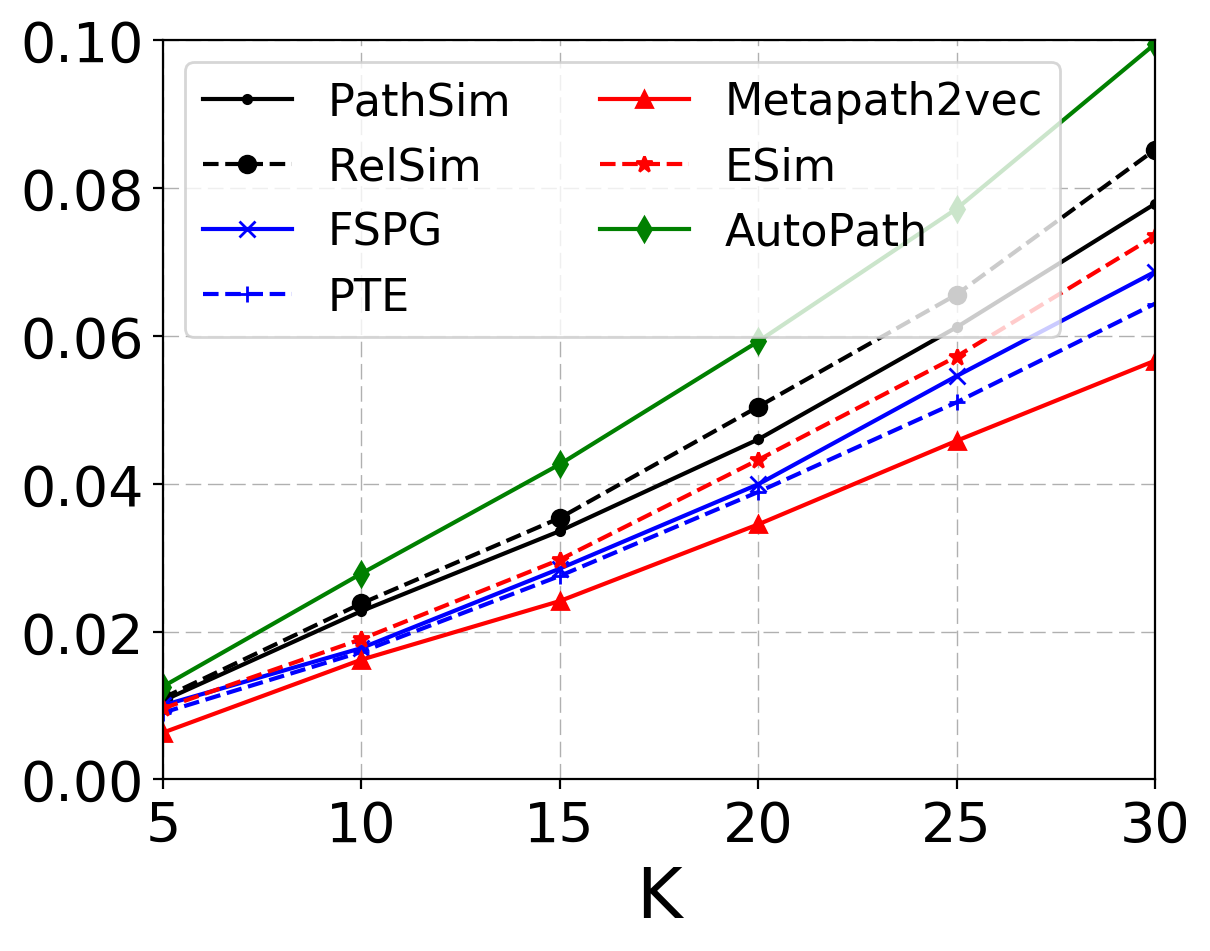}}
\hspace{-10pt}
\subfigure[Recall-DBLP]{
\includegraphics[width=0.33\textwidth]{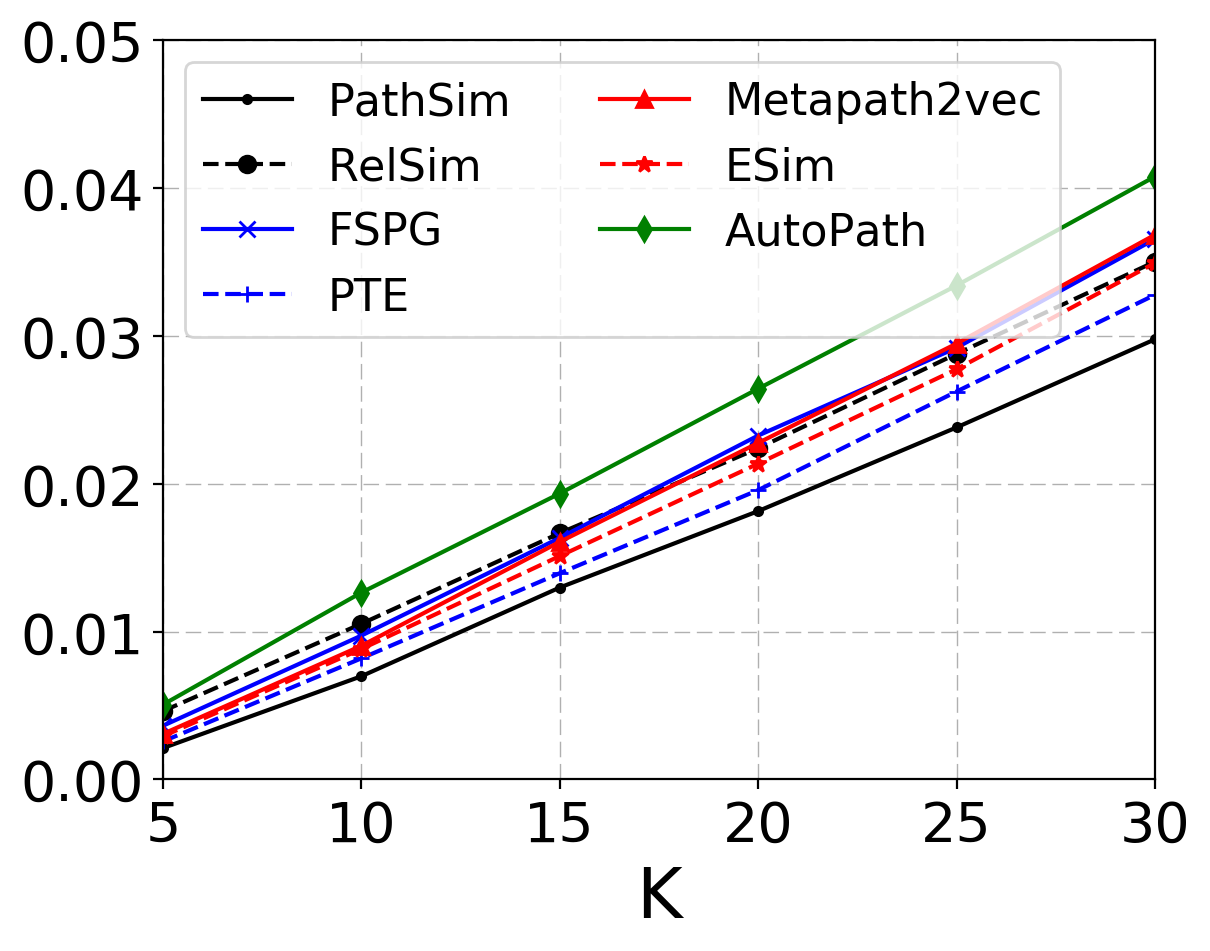}}
\hspace{-10pt}
\subfigure[Recall-Yelp]{
\includegraphics[width=0.33\textwidth]{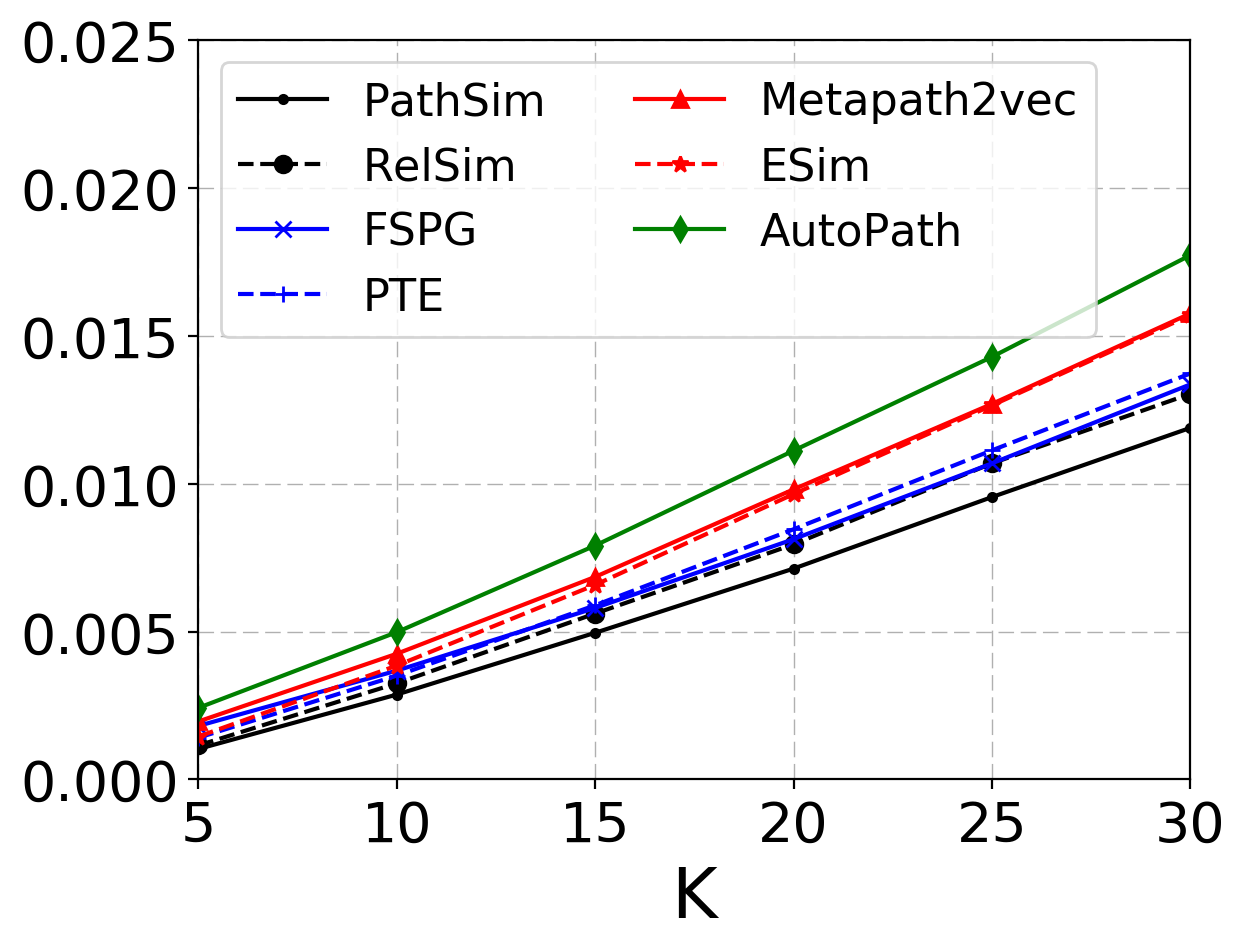}}
\caption{Quantitative evaluation results: \textit{Precision} and \textit{recall} of compared algorithms.}
\label{fig:quantity}
\end{figure}

As we can observe from Figure \ref{fig:quantity} and Table \ref{tab:metric}: (1) the compared algorithms have varying results, while \textsc{AutoPath} is able to constantly outperform all of them with significant margins on all experimented datasets, demonstrating its general and robust advantages; (2) the performance improvements of \textsc{AutoPath} are more significant on DBLP and Yelp datasets where rich node contents are available, indicating the advantage of content embedding; (3) \textit{FSPG} and \textit{RelSim} perform much better than \textit{PathSim}, and even better than the advanced network embedding algorithms, especially on DBLP, probably because they consider different weights of meta-paths. \textsc{AutoPath} also performs well on DBLP, indicating the advantage of reinforcement learning in automatically discovering important paths; (4) the runtimes of \textsc{AutoPath} are shorter than \textit{FSPG} and \textit{RelSim}, which try to enumerate or search for all useful meta-paths, especially on large networks like DBLP and Yelp, indicating its efficiency and scalability. Due to space limit, we put more discussions into our \textit{Supplementary Materials} and defer more detailed experimental studies into the future work.

\begin{table}[h!]
\centering
 \begin{tabular}{|c||c|c|c||c|c|c|}
   \hline
\multirow{2}{*}{\bf Algorithm} &\multicolumn{3}{c||}{\bf AUC} & \multicolumn{3}{c|}{\bf Runtime}\\
  \cline{2-7}
& {\bf IMDb} & {\bf DBLP} & {\bf Yelp} & {\bf IMDb} & {\bf DBLP} &{\bf Yelp}  \\
  \hline
  \hline
{\bf PathSim}& $0.584 \pm 0.018$ & $0.692 \pm 0.022$ & $0.541 \pm 0.006$ & 119s & 241s & 468s\\
\hline
{\bf RelSim}& $0.602 \pm 0.023$ & $0.788 \pm 0.028$ & $0.595 \pm 0.011$ & 325s & 1498s & 4394s\\
\hline
{\bf FSPG}& $0.568 \pm 0.011$ & $0.759 \pm 0.024$ & $0.612 \pm 0.013$ & 186s & 1062 & 3186s\\
\hline
{\bf PTE}& $0.544 \pm 0.008$ & $0.707 \pm 0.018$ & $0.608 \pm 0.015$ & 46s & 238s & 424s \\
\hline
{\bf Metapath2vec}& $0.539 \pm 0.010$ & $0.726 \pm 0.021$ & $0.622 \pm 0.015$ & 127s & 1170s & 2824s \\
\hline
{\bf ESim}& $0.573 \pm 0.012$ & $0.715 \pm 0.016$ & $0.636 \pm 0.018$ & 256s & 312s & 684s \\
\hline
{\bf AutoPath}& $\mathbf{0.635 \pm 0.015}$ & $\mathbf{0.840 \pm 0.018}$ & $\mathbf{0.713 \pm 0.016}$ & 163s & 466s & 1620s \\
\hline
 \end{tabular}
 \vspace{10pt}
 \caption{ \label{tab:metric}Quantitative evaluation results: \textit{AUC} and \textit{runtime} of compared algorithms.}
\end{table}

\subsection{Qualitative Analysis}
As we stress in this work, a unique advantage of \textsc{AutoPath} is the automatic discovery of useful meta-paths from enormous search spaces without a pre-defined maximum length. To demonstrate such utility, after training our model, we plan on random nodes for 10,000 times and summarize the most frequently traveled meta-paths in Table \ref{tab:path}. As we can see, the meta-paths with variable lengths and importance discovered by our algorithm are indeed intuitive for each dataset, indicating the power of it in automatically discovering important paths.

\begin{table}[h!]
\centering
 \vspace{10pt}
\small
 \begin{tabular}{|cc|cc|cc|}
   \hline
\multicolumn{2}{|c|}{\bf IMDb} &\multicolumn{2}{c|}{\bf DBLP} &\multicolumn{2}{c|} {\bf Yelp}\\
  \hline
  \hline
M -- A -- M &({\it 0.372}) & A -- P -- V -- P -- A &({\it 0.781})& B -- L -- B& ({\it 0.414})\\
  \hline
M -- D -- M &({\it 0.315}) & A -- P -- A &({\it 0.132}) & B -- U -- B& ({\it 0.277})\\
  \hline
M -- U -- M &({\it 0.298}) & A -- P -- A -- P -- A &({\it 0.046}) & B -- S -- B &({\it 0.236})\\
  \hline
 \end{tabular}
  \vspace{10pt}
 \caption{ \label{tab:path}Top 3 meta-paths automatically found and deemed important by \textsc{AutoPath}.}
\end{table}

%% file: sec-con.tex
\section{Conclusions}
\label{sec:con}
Heterogeneous networks have been intensively studied recently, due to its power of incorporating different types of data from various sources. 
In this work, we focus on the key challenge of learning with heterogeneous networks, \ie, similarity modeling.
To fully leverage both structural and content information over heterogeneous networks, we break free the requirement of pre-defined meta-paths through automatic path discovery with efficient reinforcement learning and incorporate rich node contents to empower discriminative path exploration through deep content embedding. 
We demonstrate the effectiveness and efficiency of our \textsc{AutoPath} algorithm through extensive quantitative and qualitative experiments on three large-scale real-world heterogeneous networks.

For future works, more in-depth experiments can be done to study the individual effectiveness of our reinforcement learning and content embedding frameworks. Meanwhile, various improvements can also be thought of for both of them, such as the embedding of more complex contents like texts and images, the interpretation of discovered paths, and the generation of heterogeneous network embedding for various other downstream applications.

%% file: sec-ack.tex
\section*{Acknowledgement}
Research was sponsored in part by U.S. Army Research Lab. under Cooperative Agreement No. W911NF-09-2-0053 (NSCTA), DARPA under Agreement No. W911NF-17-C-0099, National Science Foundation IIS 16-18481, IIS 17-04532, and IIS-17-41317, DTRA HDTRA11810026, and grant 1U54GM114838 awarded by NIGMS through funds provided by the trans-NIH Big Data to Knowledge (BD2K) initiative (www.bd2k.nih.gov). 


%% file: autopath.bbl
\begin{thebibliography}{10}
\providecommand{\url}[1]{\texttt{#1}}
\providecommand{\urlprefix}{URL }

\bibitem{bello2016neural}
Bello, I., Pham, H., Le, Q.V., Norouzi, M., Bengio, S.: Neural combinatorial
  optimization with reinforcement learning. In: ICLR (2017)

\bibitem{das2017go}
Das, R., Dhuliawala, S., Zaheer, M., Vilnis, L., Durugkar, I., Krishnamurthy,
  A., Smola, A., McCallum, A.: Go for a walk and arrive at the answer:
  Reasoning over paths in knowledge bases using reinforcement learning. In:
  ICLR (2018)

\bibitem{dong2017metapath2vec}
Dong, Y., Chawla, N.V., Swami, A.: metapath2vec: Scalable representation
  learning for heterogeneous networks. In: KDD. pp. 135--144 (2017)

\bibitem{fang2016semantic}
Fang, Y., Lin, W., Zheng, V.W., Wu, M., Chang, K., Li, X.L.: Semantic proximity
  search on graphs with metagraph-based learning. In: ICDE. pp. 277--288 (2016)

\bibitem{fu2017hin2vec}
Fu, T.y., Lee, W.C., Lei, Z.: Hin2vec: Explore meta-paths in heterogeneous
  information networks for representation learning. In: CIKM. pp. 1797--1806
  (2017)

\bibitem{han2011data}
Han, J., Pei, J., Kamber, M.: Data mining: concepts and techniques. Elsevier
  (2011)

\bibitem{harper2016movielens}
Harper, F.M., Konstan, J.A.: The movielens datasets: History and context. TIIS
  5(4), ~19 (2016)

\bibitem{huang2017heterogeneous}
Huang, Z., Mamoulis, N.: Heterogeneous information network embedding for meta
  path based proximity. arXiv preprint arXiv:1701.05291  (2017)

\bibitem{khalil2017learning}
Khalil, E., Dai, H., Zhang, Y., Dilkina, B., Song, L.: Learning combinatorial
  optimization algorithms over graphs. In: NIPS. pp. 6351--6361 (2017)

\bibitem{kingma2014adam}
Kingma, D.P., Ba, J.: Adam: A method for stochastic optimization. In: ICLR
  (2015)

\bibitem{le2013building}
Le, Q.V.: Building high-level features using large scale unsupervised learning.
  In: ICASSP. pp. 8595--8598 (2013)

\bibitem{lillicrap2015continuous}
Lillicrap, T.P., Hunt, J.J., Pritzel, A., Heess, N., Erez, T., Tassa, Y.,
  Silver, D., Wierstra, D.: Continuous control with deep reinforcement
  learning. arXiv  (2015)

\bibitem{liu2017semantic}
Liu, Z., Zheng, V.W., Zhao, Z., Zhu, F., Chang, K., Wu, M., Ying, J.: Semantic
  proximity search on heterogeneous graph by proximity embedding. In: AAAI. pp.
  154--160 (2017)

\bibitem{meng2015discovering}
Meng, C., Cheng, R., Maniu, S., Senellart, P., Zhang, W.: Discovering
  meta-paths in large heterogeneous information networks. In: WWW. pp. 754--764
  (2015)

\bibitem{mnih2015human}
Mnih, V., Kavukcuoglu, K., Silver, D., Rusu, A.A., Veness, J., Bellemare, M.G.,
  Graves, A., Riedmiller, M., Fidjeland, A.K., Ostrovski, G., et~al.:
  Human-level control through deep reinforcement learning. Nature  518(7540),
  529 (2015)

\bibitem{pennington2014glove}
Pennington, J., Socher, R., Manning, C.D.: Glove: Global vectors for word
  representation. In: EMNLP. pp. 1532--1543 (2014)

\bibitem{schulman2015trust}
Schulman, J., Levine, S., Abbeel, P., Jordan, M., Moritz, P.: Trust region
  policy optimization. In: ICML. pp. 1889--1897 (2015)

\bibitem{shang2016meta}
Shang, J., Qu, M., Liu, J., Kaplan, L.M., Han, J., Peng, J.: Meta-path guided
  embedding for similarity search in large-scale heterogeneous information
  networks. arXiv preprint arXiv:1610.09769  (2016)

\bibitem{shi2017prep}
Shi, Y., Chan, P.W., Zhuang, H., Gui, H., Han, J.: Prep: Path-based relevance
  from a probabilistic perspective in heterogeneous information networks. In:
  KDD. pp. 425--434 (2017)

\bibitem{shi2018aspem}
Shi, Y., Gui, H., Zhu, Q., Kaplan, L., Han, J.: Aspem: Embedding learning by
  aspects in heterogeneous information networks. In: SDM (2018)

\bibitem{sun2012mining}
Sun, Y., Han, J.: Mining heterogeneous information networks: principles and
  methodologies. Synthesis Lectures on Data Mining and Knowledge Discovery
  3(2),  1--159 (2012)

\bibitem{sun2011pathsim}
Sun, Y., Han, J., Yan, X., Yu, P.S., Wu, T.: Pathsim: Meta path-based top-k
  similarity search in heterogeneous information networks. VLDB  4(11),
  992--1003 (2011)

\bibitem{sutton2000policy}
Sutton, R.S., McAllester, D.A., Singh, S.P., Mansour, Y.: Policy gradient
  methods for reinforcement learning with function approximation. In: NIPS. pp.
  1057--1063 (2000)

\bibitem{tang2015pte}
Tang, J., Qu, M., Mei, Q.: Pte: Predictive text embedding through large-scale
  heterogeneous text networks. In: KDD. pp. 1165--1174 (2015)

\bibitem{tang2008arnetminer}
Tang, J., Zhang, J., Yao, L., Li, J., Zhang, L., Su, Z.: Arnetminer: extraction
  and mining of academic social networks. In: KDD. pp. 990--998 (2008)

\bibitem{wan2015graph}
Wan, M., Ouyang, Y., Kaplan, L., Han, J.: Graph regularized meta-path based
  transductive regression in heterogeneous information network. In: SDM. pp.
  918--926 (2015)

\bibitem{wang2015knowsim}
Wang, C., Song, Y., Li, H., Zhang, M., Han, J.: Knowsim: A document similarity
  measure on structured heterogeneous information networks. In: ICDM. pp.
  1015--1020 (2015)

\bibitem{wang2016relsim}
Wang, C., Sun, Y., Song, Y., Han, J., Song, Y., Wang, L., Zhang, M.: Relsim:
  relation similarity search in schema-rich heterogeneous information networks.
  In: SDM. pp. 621--629 (2016)

\bibitem{xiong2017deeppath}
Xiong, W., Hoang, T., Wang, W.Y.: Deeppath: A reinforcement learning method for
  knowledge graph reasoning. In: EMNLP (2017)

\bibitem{yang2017bridging}
Yang, C., Bai, L., Zhang, C., Yuan, Q., Han, J.: Bridging collaborative
  filtering and semi-supervised learning: A neural approach for poi
  recommendation. In: KDD. pp. 1245--1254 (2017)

\bibitem{yang2018did}
Yang, C., Zhang, C., Chen, X., Ye, J., Han, J.: Did you enjoy the ride:
  Understanding passenger experience via heterogeneous network embedding. In:
  ICDE (2018)

\bibitem{yang2017bi}
Yang, C., Zhong, L., Li, L.J., Jie, L.: Bi-directional joint inference for user
  links and attributes on large social graphs. In: WWW. pp. 564--573 (2017)

\bibitem{zhao2017meta}
Zhao, H., Yao, Q., Li, J., Song, Y., Lee, D.L.: Meta-graph based recommendation
  fusion over heterogeneous information networks. In: KDD. pp. 635--644 (2017)

\end{thebibliography}
